\documentclass{epl}

\usepackage{graphicx}

\title{Soliton ratchets out of point-like inhomogeneities}
\author{Luis Morales-Molina\inst{1,2} \and Franz G.\ Mertens\inst{1}
\and Angel S\'anchez\inst{2}}
\shortauthor{L.\ Morales-Molina {\em et al.}}
\institute{
  \inst{1} Physikalisches Institut, Universit\"at Bayreuth, D-85440 Bayreuth,
Germany\\
  \inst{2} Grupo Interdisciplinar de Sistemas Complejos
(GISC) and
Departamento de\\ Ma\-te\-m\'a\-ti\-cas,
Universidad Carlos III de Madrid, Avenida de la Universidad 30,\\ 28911
Legan\'es, Madrid, Spain
}

\begin{document}

\maketitle

\begin{abstract}
We introduce and study a novel design for a ratchet potential for soliton
excitations. The potential is implemented by means of an array of point-like
(delta)
inhomogeneities in an otherwise homogeneous potential. We develop a collective 
coordinate theory that predicts that the effective potential acting on the
soliton is periodic but asymmetric and gives rise to the ratchet effect. 
Numerical simulations fully confirm this prediction; quantitative agreement
is reached by an improved version of the theory.
Although we specifically show that it 
is most interesting for building Josephson junction ratchets
capable to rectify time-symmetric ac forces,
the proposed
mechanism is very general and can appear in many contexts, including
biological systems.
\end{abstract}

\section{Introduction}

Ratchet systems have been the object of intensive studies due to their
promising applications in biological systems\cite{Maddox,Reviews} and micro-
and nano-technologies \cite{Linke,Grifoni}.
Recently, a great deal of effort has been devoted to generalize the ratchet
mechanism for point particles to spatially extended systems \cite{extended}.
One proposal along this direction relates to the existence of net transport in homogeneous
extended systems driven by homogeneous ac forces \cite{flach,salerno}.
However, although this is a very good and feasible way to induce transport
with specified properties, it cannot be used to rectify time-symmetric forces, because it relies on the breaking of the time symmetry of the ac force. 
In case the ac force is symmetric, the alternative route to rectification is
to introduce spatial inhomogeneity. Models in this class
have been studied theoretically \cite{falo} and
also implemented in superconducting devices\cite{falo2,Grifoni}.
In this case, a drawback 
is the difficulty of their fabrication at the
micro- or nano-scale, because controlling the asymmetry is very complex 
\cite{Grifoni}.  An additional factor that has to be considered in 
this context is 
the interplay between disorder and nonlinearity, which can be of fundamental
relevance in the design of these new transport devices and particularly when
competition of scales takes place\cite{angel0,angelbis}.

In this letter, we present a much simpler design for an extended ratchet that works irrespective of the symmetry of the ac force. Specifically, we focus on the sine-Gordon (sG) 
model, among other reasons because of its important applications to
superconducting devices such as long Josephson Junctions (LJJ). In this 
context, our proposal is based on the inclusion of point-like inhomogeneities,
which correspond to micro-shorts along the LJJ \cite{mclaughlin,Ustinov}.
Notwithstanding, the mechanism is very general and it can be applied to
many other 
soliton-bearing systems where the interaction of kinks with point-like
inhomogeneities is similar to that occurring in the sG model 
\cite{kivshar,angel,angel2}. 
We begin by introducing the perturbed sG model, driven by an ac force in
the presence of a periodic array of inhomogeneities. We first
study analytically this 
problem by means of a collective coordinate for the motion of the kink
center.
We thus show that under certain conditions the effective
kink dynamics
is that of a particle ratchet.
We then study, by numerical simulations of the full system, the behavior
of the mean velocity as a function of the driving amplitude for 
different frequencies. Although qualitative agreement with the analytical
predictions is found, the quantitative comparison is not so good. 
We subsequently improve our theory by introducing the width degree 
of freedom, whose quantitative success makes
clear the physical mechanism needed for a correct
description of the phenomenon.
We conclude by summarizing our main results and pointing out future
research along these lines.

\section{Model}

Our starting point is the following perturbed equation:
\begin{equation}
\ \phi_{tt}+\beta\phi_{t}-\phi_{xx}+ [1+V(x)]\sin\phi=A\sin(\omega t+\delta_{0}),  \label{1}
\end{equation}
where the term $V(x) \sin(\phi)$ accounts for local inhomogeneities
which can
be, for instance, variations of the critical current in a LJJ,
and 
$A\sin(\omega t+\delta_{0})\equiv f(t)$ corresponds to an ac bias current
\cite{carapella}.
The kink solution of Eq.\ (\ref{1}) in the absence of perturbations, i.e., 
$\beta=V=A=0$, is given by the expression
\begin{equation}\label{2}
\phi(x,t)=4\arctan\left(\exp\left[\frac{x-X(t)}{l_{0}\sqrt{1-v^{2}}}\right]\right),
\end{equation}
where
$X(t)$ and
$v=\dot{X}(t)$
are the kink 
position and velocity, respectively, and
$l_{0}$ is a measure of the kink width at rest. For the sine-Gordon case 
$l_0=1$, but we choose to leave this as an explicit parameter to exhibit
the physical relevance of this magnitude; besides, in
many other soliton-bearing systems collective coordinate equations
are the same 
as in this case but 
$l_0\neq 1$ (e.g., the $\phi^4$ system).
The solution in Eq.\ (\ref{2}) represents, always in the
context of LJJ, a flux quantum (fluxon) propagating along the junction. 

The question now is: What is the ideal shape of the function $V(x)$
to turn our system into a ratchet device based on the fluxon?
It is important to recall that, when driven by symmetric ac forces,
a damped fluxon can only exhibit oscillatory motion \cite{niurka3}.
This problem was overcome in
previous works, and soliton ratchet behavior has been found for this
system when the potential of the unperturbed sine-Gordon equation becomes 
asymmetric \cite{salerno2}, or when the system is under an
inhomogeneous magnetic field \cite{carapella}.
To our knowledge, none of
the previous works resorted to inhomogeneities 
for breaking the symmetry of the system, as we now do.
Our proposal consists of 
an array of point-like inhomogeneities, which in the case of LJJ
can be modelled as delta functions if their length is less than
the Josephson penetration length.
%In this case $V(x)$ becomes
%\begin{equation}\label{4}
%\ V(x)=\epsilon \sum_{n}\delta(x-x_{n})
%\end{equation}
%where $\delta(x-x_{n})$ 
%with $n=1,2,..$ are delta functions located at the points $x_{n}$.\\
We will choose $\{x_{n}\}$ to form a periodic, asymmetric
array; in particular, for this work we have specifically chosen 
three inhomogeneities per spatial period,
\begin{equation}\label{5}
\ V(x)=\epsilon \sum_{n}\left[\delta(x-x_{1}-nL)+\delta(x-x_{2}-nL)+\delta(x-x_{3}-nL)\right],
\end{equation}
where the parameters should satisfy the constraints 
$a,b,c\sim l_{0}$;  $a,b<c$  with $a\neq b$, 
where  $L=a+b+c$, $a=x_{2}-x_{1}$, $b=x_{3}-x_{2}$ and $c=L-x_{3}$.
As will be shown below, 
for a system satisfying these conditions we obtain net motion with
a behavior that resembles very closely the one found
in ratchet-like systems for point-particles. We have to stress that
the distances between the delta functions have to be 
of the same length scale as the kink width; otherwise, different
behaviors could arise like those demonstrated in \cite{angel2}, as the 
interference between adjacent deltas is lost.
Within that requirement, our proposal is very versatile, as 
it is possible in principle to induce directional motion by using an array
whose configuration presents more than three inhomogeneities per period,
in case a different ratchet potential were required. 

\section{Collective Coordinate Theory}

As a first step to justify our choice for the perturbative term $V(x)$, 
we present a simple collective coordinate analysis of its effect on the
soliton dynamics. The idea of this well-known approximate technique for
treating soliton-bearing equations is to assume that perturbations affect
mostly the motion of the soliton center (and/or other parameters, as we
will see below) and to proceed to a drastic reduction of the number of 
degrees of freedom by deriving an effective equation for the corresponding
collective coordinate (see, e.g., \cite{angelbis} for a recent review
and further references). One of the 
easiest procedures to derive equations for the collective
coordinate is by means of the conservations laws,
making use of the so-called adiabatic approach, first proposed by 
McLaughlin and Scott \cite{mclaughlin}. Following straightforwardly 
the procedure in this reference, it is a matter of algebra to show,
using Eq.\ (\ref{2}) 
as an {\em Ansatz} for the perturbed Eq.\ (\ref{1}),
that the corresponding equation of motion for the kink center in 
the limit of small velocities $v^2\ll 1$ is 
\begin{equation}\label{6}
M_{0}\ddot{X}+ \beta M_{0}\dot{X}=-\frac{\partial
U}{\partial X}-q A\sin(\omega t+\delta_{0}),
\end{equation}
where
\begin{equation}\label{7}
U(X)=2\epsilon \sum_{n}\left[\frac{1}{\cosh^2(X-x_{1}-nL)}+
\frac{1}{\cosh^2(X-x_{2}-nL)}+\frac{1}{\cosh^2(X-x_{3}-nL)}\right],
\end{equation}
is the effective potential function, we have for this one time 
put $l_0=1$, and
$M_{0}=8$ and $q=2\pi$ are the soliton rest mass and 
topological charge, respectively. 

\begin{figure}
\begin{center}
\includegraphics[width=0.49\textwidth]{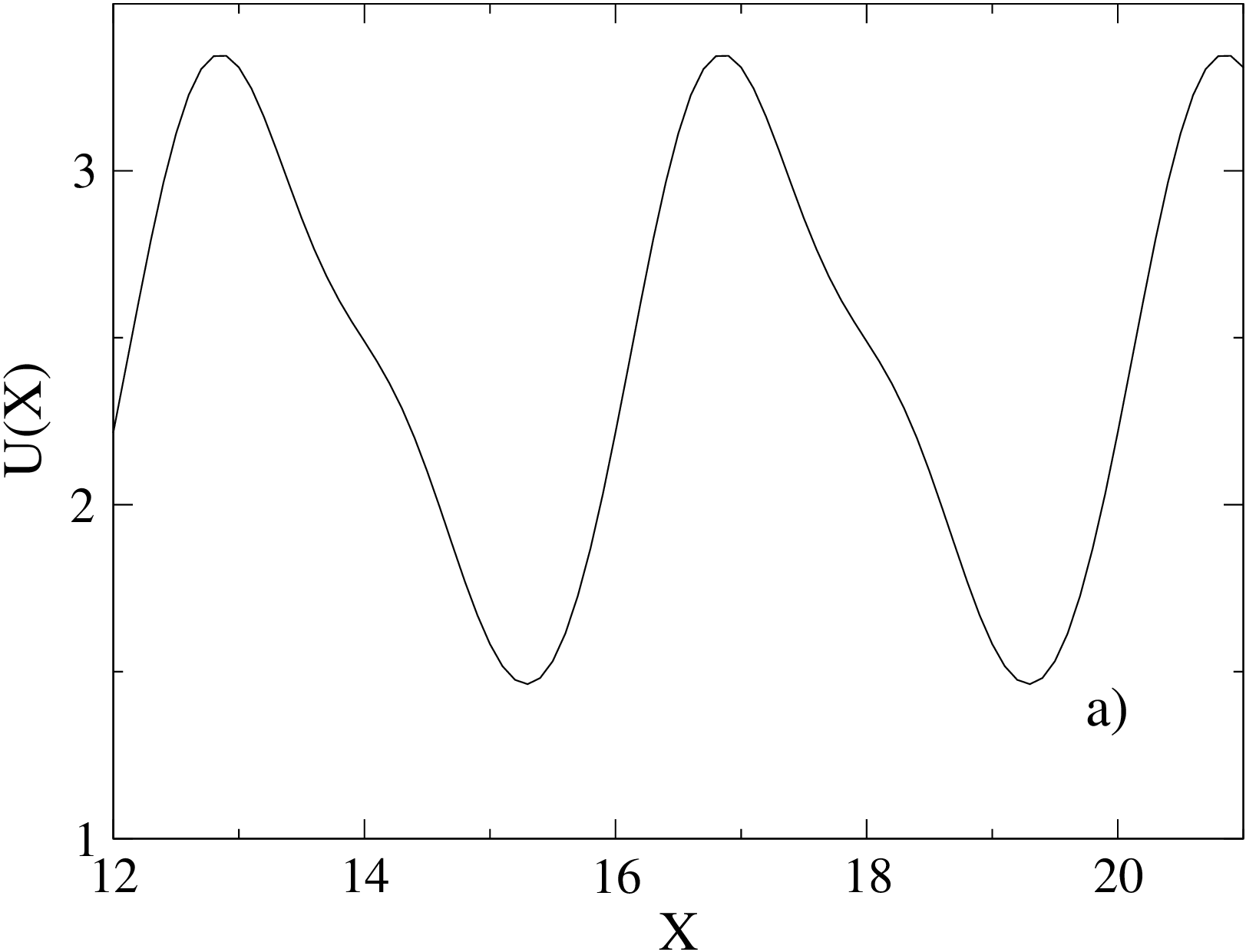}
\includegraphics[width=0.49\textwidth]{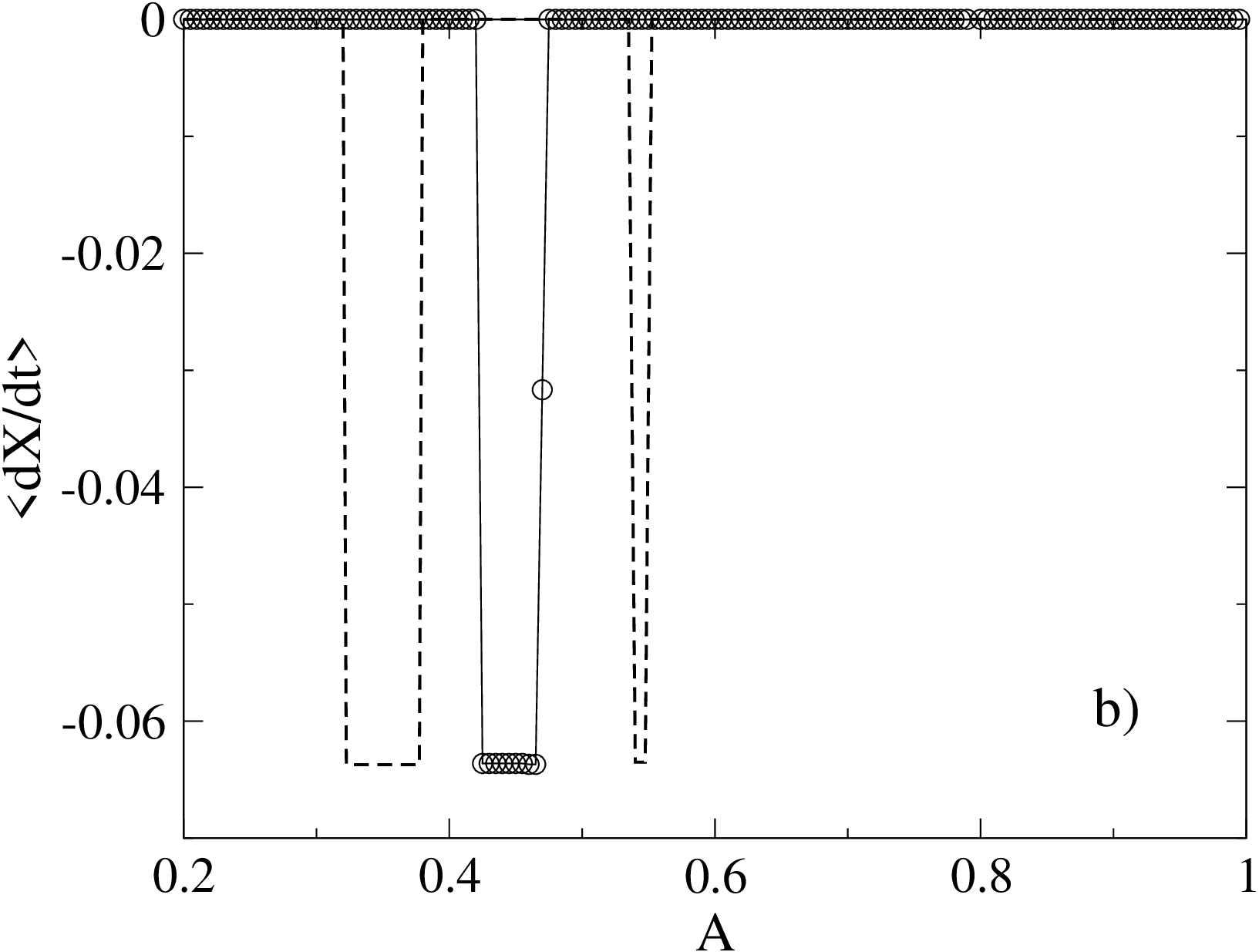}
\caption{a) Effective potential for the kink center within the adiabatic 
approach, originating in the perturbation $V(x)$ defined in Eq.\ (\ref{5})
with $\epsilon=0.8$, $x_{1}=0.5$,
$x_{2}=1.$, $x_{3}=2.3$ and period $L=4$. b) Mean kink velocity $\langle dX/dt\rangle$ vs driving amplitude $A$ for the frequency $\omega=0.1$. Circles: direct numerical simulation of Eq.\ (\ref{1}), the
line being only a guide to the eye; 
dashed line: collective coordinate approach, Eq.\ (\ref{6}).}
\label{ratchet1}
\end{center}
\end{figure}

The potential function given by Eq.\ (\ref{7}) is depicted in 
Fig.\ \ref{ratchet1}a) for the perturbation $V(x)$ defined in Eq.\ (\ref{5}),
with 
$x_{1}=0.5$,
$x_{2}=1.$, $x_{3}=2.3$ and period $L=4$.
As can be seen from the figure, it corresponds 
to an asymmetric potential characteristic of ratchet systems; in fact, equation
(\ref{6}) is the same as that for a point particle in a rocking ratchet 
(see \cite{Reviews}). We stress, however, that Eq.\ (\ref{6}), as it is,
describes an {\em inertial} rocking ratchet, similar to those studied
in \cite{Jung}. As in this case the dynamics is much more complicated, 
involving dependencies on the initial conditions and on other factors, 
in the following we restrict ourselves to the overdamped case (the 
common situation in ratchet systems \cite{Reviews}) by taking $\beta=1$.
As an immediate consequence, within this approximation,
we can expect that the soliton center should move towards the left.
Fig.\ \ref{ratchet1}b) shows, for one particular value of the frequency,
the prediction of Eq.\ (\ref{6}) in this overdamped approach, 
confirming this expectation and showing the typical window 
behavior of ratchets (see below).

\section{Numerical results}

To check the prediction of the simple theory we have summarized above,
we have carried out numerical simulations of Eq.\ (\ref{1}). We have
used a Strauss-V\'azquez numerical scheme \cite{sv} with free boundary
conditions. The spatial and temporal integration steps were 
$\Delta x=0.1$ and $\Delta t=0.01$ respectively. The spatial interval
for the simulations was $[-50,150]$, with the inhomogeneities arranged
periodically according to our three delta unit cell in $[0,100]$. The
system was simulated up to times as long as $T=4000$ time units. 
Finally, the numerical representation for delta functions
is $1/\Delta x$ as usual \cite{kivshar,angel2}.

\begin{figure}
\begin{center}
\includegraphics[width=0.49\textwidth]{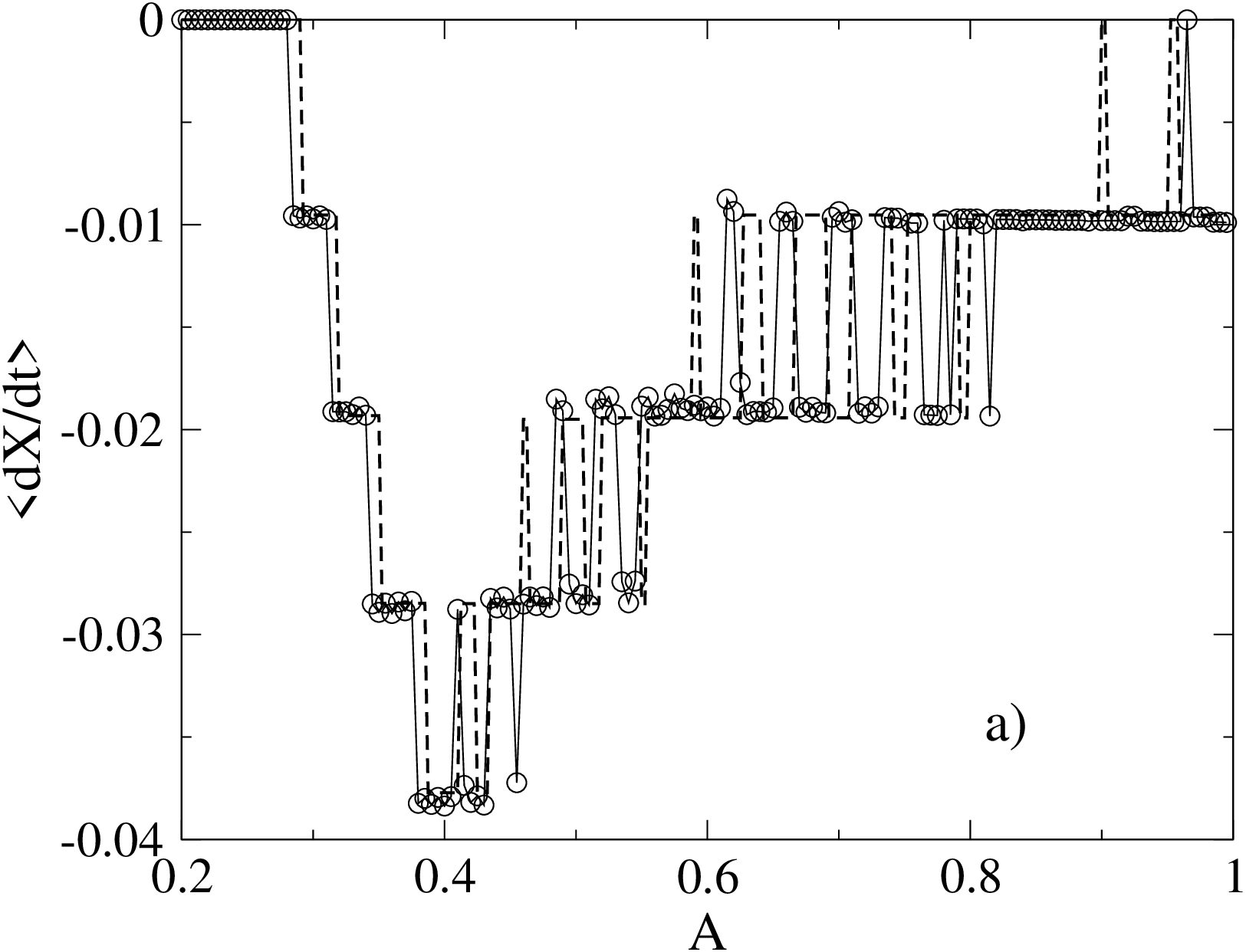}
\includegraphics[width=0.49\textwidth]{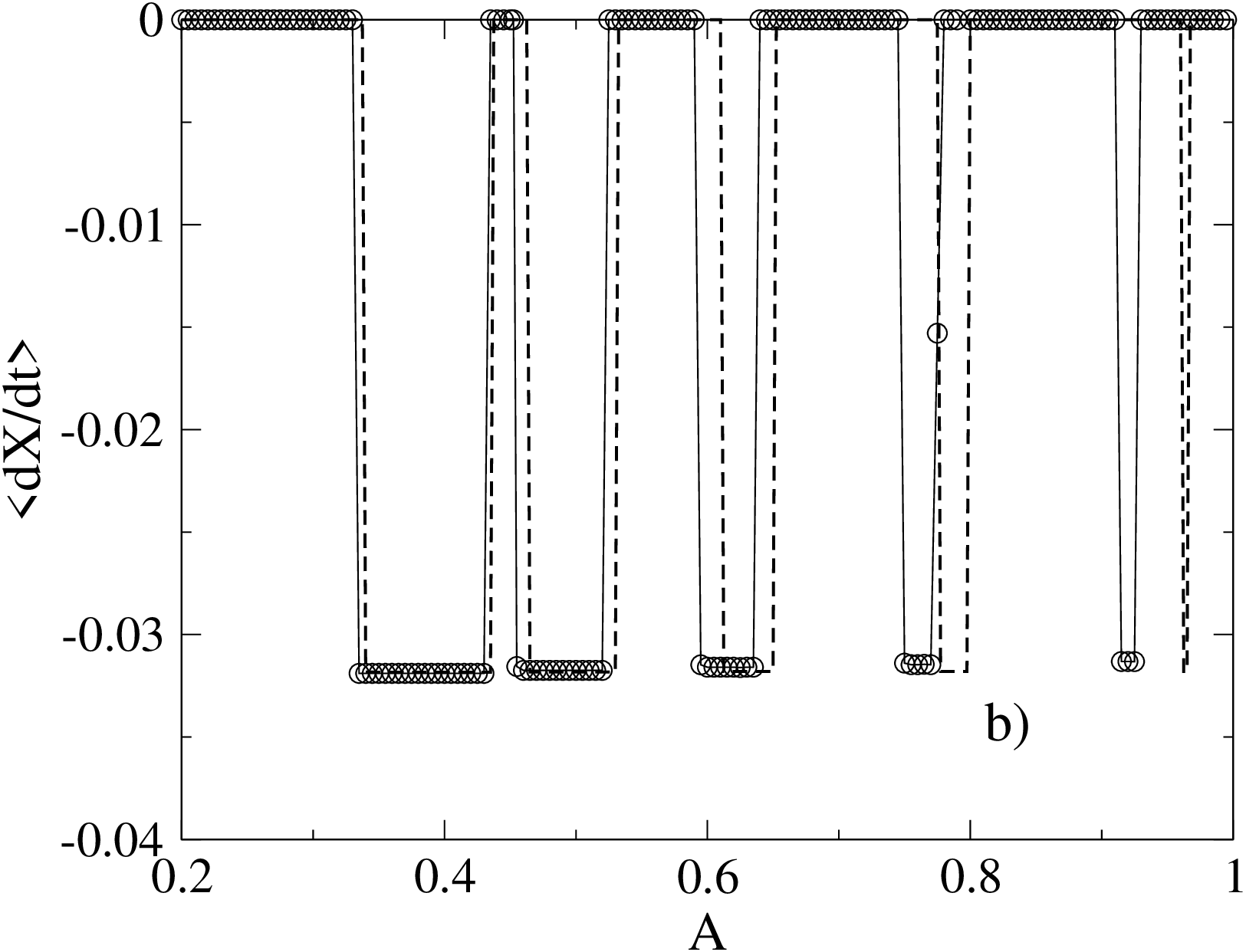}

\vspace*{5mm}

\includegraphics[width=0.49\textwidth]{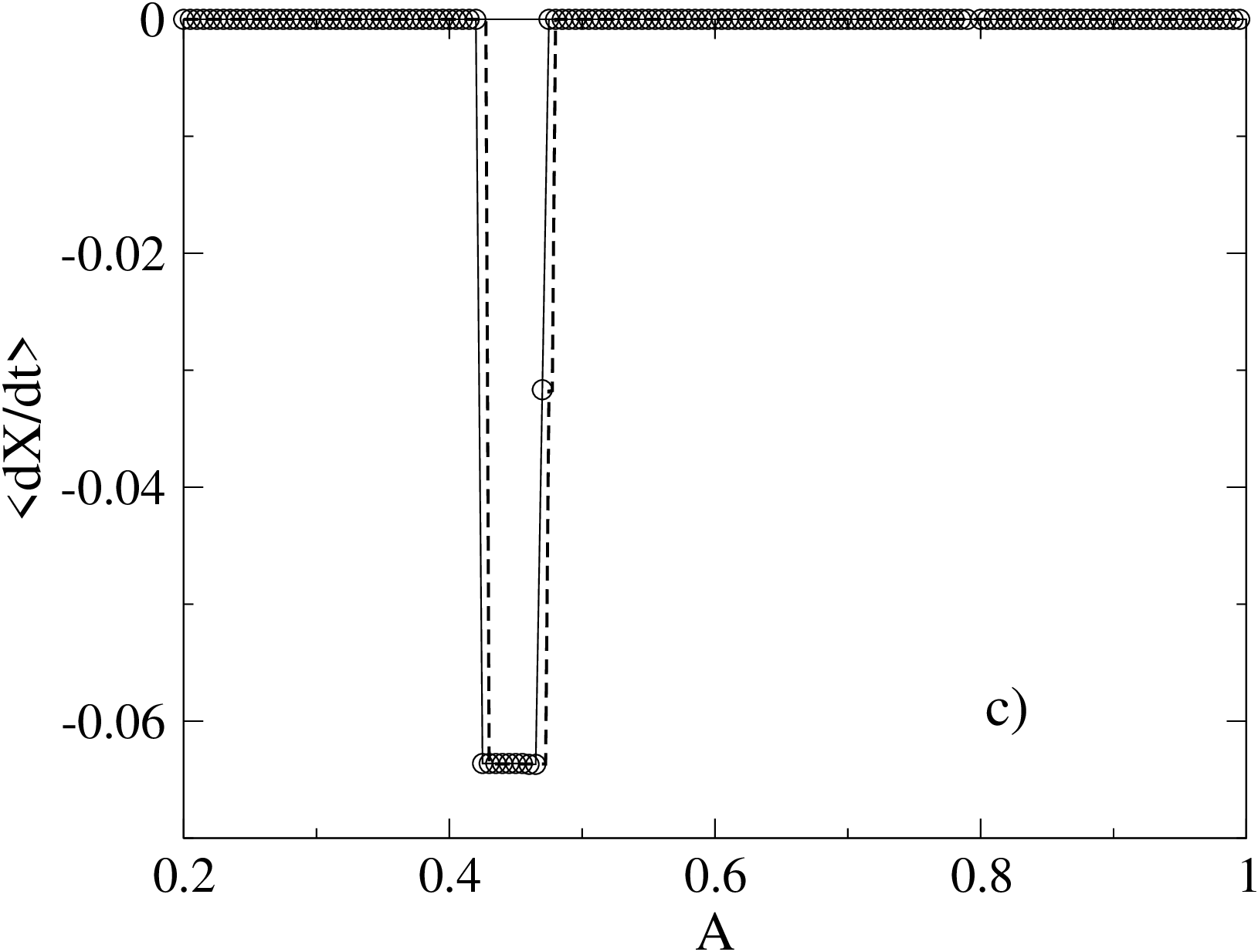}
\caption{Mean kink velocity $\langle dX/dt\rangle$ vs driving amplitude $A$ for different 
frequencies: a) $\omega=0.015$, b) $\omega=0.05$, c) $\omega=0.1$. 
Other parameters are the same as in Fig.\ \ref{ratchet1}.
Circles: direct numerical simulation of Eq.\ (\ref{1}), the
line being only a guide to the eye; 
dashed line: improved collective coordinate theory, Eqs.\ (\ref{15}),
(\ref{15b}).} 
\label{f.2}
\end{center}
\end{figure}

The simulation results are summarized in Fig.\ \ref{f.2}. As we may see, the 
system behaves very much like a point particle ratchet (cf., e.g., 
\cite{sironis,mag}; see also Fig.\ 5 in \cite{falo2}).
Indeed, we appreciate the existence of windows of motion
separated by gaps where the motion is oscillatory with mean velocity zero
and
whose extension increases upon increasing frequency.
The explanation of these gaps and the observation of ``quantized''
velocity values as typical signatures of ratchet behavior can be 
found in \cite{nueva} for the particle-like case and in \cite{falo2}
for the extended system.
We have also
verified that, when the soliton leaves the zone in which the 
array is contained, its motion becomes purely oscillatory,
as expected \cite{niurka3}.

Turning now to a more detailed comparison, we have to admit that the
agreement with the collective coordinate theory of the previous 
section is not quite satisfactory.  
Fig.\ \ref{ratchet1}b) makes this point clear by showing that neither
the number of windows nor, of course, their locations, are correctly 
predicted, even in the simpler low frequency case.
Searching for an explanation of this problem, we analyzed in depth
the simulations, finding out that a possible
reason for this discrepancy is 
that the soliton shape changes during its motion along the 
inhomogeneities array (namely, its width is oscillating). This 
feature can not be accounted for within
the framework of our theory above and therefore we set out to 
improve it in the next section. 

\section{Improved Collective Coordinate Theory}

In order to account for the phenomenology observed in the simulations, we
resort to the  {\em generalized travelling wave ansatz} for solving our problem,
first proposed in \cite{mertens,niurka2} for one and two degrees of freedom.
As our starting point we rewrite Eq.\ (\ref{1}) in the following way
\begin{equation}\label{10}
 \dot{\phi}=\frac{\delta H}{\delta \psi},\>\>\>
\dot{\psi}=-\frac{\delta H}{\delta \phi}-\beta\dot{\phi}
-\sin(\phi)V(x)+f(t),
\end{equation} 
where $\psi=\dot{\phi}$ and $H$ is the Hamiltonian for the unperturbed
problem. 
Following the procedure as in \cite{niurka2} and assuming that
$\phi$ has the form of the so-called Rice {\em Ansatz} \cite{rice}
\begin{equation}\label{11}
\phi(x,t)=\phi_{K}[x-X(t),l(t)]=4\arctan\left(\exp\left[\frac{x-X(t)}{l(t)}\right]\right),
\end{equation} 
\enlargethispage{3mm}
where $l(t)$ is intended to account
for the observed oscillations of the kink width, 
we find:
%\begin{eqnarray}\label{15}
%\ M_{0}l_{0}\frac{\ddot{X}}{l}-M_{0}l_{0}\frac{\dot{X}\dot{l}}{l^2}&=&F^{stat} -\beta M_{0}l_{0}\frac{\dot{X}}{l}+F_{ext},  
%\\ \label{16}
%\alpha
%  M_{0}l_{0}\frac{\ddot{l}}{l}+M_{0}l_{0}\frac{\dot{X}^2}{l^2}&=&-\beta\alpha
%  M_{0}l_{0}\frac{\dot{l}}{l}+K^{int}(l,\dot{l},\dot{X})+K,
%\end{eqnarray}
%with the terms appearing in the right hand side being given by 
%\begin{eqnarray}
%\label{mia2}
%F_{ext}&=&\int_{-\infty}^{\infty}dx N(x,t,\psi,u(t),V(x))\frac{\partial \psi}{\partial X},  \qquad F^{stat}=-\frac{\partial E}{\partial X},\\
% K&=&\int_{-\infty}^{\infty}dx N(x,t,\psi,u(t),V(x))\frac{\partial \psi}{\partial l},\qquad K^{int}(l,\dot{l},\dot{X})=-\frac{\partial E}{\partial l},\\
%E&=&\frac{1}{2}\frac{l_{0}}{l} M_{0}\dot{X}^2+\frac{1}{2}\frac{l_{0}}{l}\alpha M_{0}\dot{l}^2+\frac{1}{2} M_{0}\left(\frac{l_{0}}{l}+\frac{l}{l_{0}}\right),
%\label{mia1}
%\end{eqnarray}
%where $M_{0}=8$, $l_{0}=1$, $\alpha=\pi^2/12$. Finally, evaluation of
%Eqs.\ (\ref{mia2}) through (\ref{mia1}) allows us to write 
%Eqs.\ (\ref{15}) and (\ref{16}) in the form

\begin{equation}\label{12}
\ M_{0}l_{0}\frac{\ddot{X}}{l}+\beta M_{0}l_{0}\frac{\dot{X}}{l}-M_{0}l_{0}
\frac{\dot{X}\dot{l}}{l^2}=-\frac{\partial U}{\partial X}
-qf(t),
\end{equation}
\begin{equation}\label{13}
\alpha
  M_{0}l_{0}\frac{\ddot{l}}{l}+\beta\alpha
  M_{0}l_{0}\frac{\dot{l}}{l}+M_{0}l_{0}\frac{\dot{X}^2}{l^2}=
K^{int}(l,\dot{l},\dot{X})-\frac{\partial U}{\partial l},
\end{equation}
where $K^{int}(l,\dot{l},\dot{X})=-\frac{\partial E}{\partial l}$ with 
\begin{equation}\label{14}
E=\frac{1}{2}\frac{l_{0}}{l} M_{0}\dot{X}^2+\frac{1}{2}\frac{l_{0}}{l}\alpha M_{0}\dot{l}^2+\frac{1}{2} M_{0}\left(\frac{l_{0}}{l}+\frac{l}{l_{0}}\right),
\end{equation}
$M_{0}=8$, $l_{0}=1$, $\alpha=\pi^2/12$, $q=2\pi$ and $U(X,l)$ has
the same form as $U(X)$ 
in Eq.\ (\ref{7}) but with the denominators recast as
$\cosh^2(X-x_{i}-nL)/l$, $i=1,2,3$.

In order to make more transparent the physical meaning of these equations, 
we can change variables by introducing
the momentum $P(t)=M_{0}l_{0}\dot{X}/l(t)$. Our equations become
\begin{eqnarray}\label{15}
\frac{dP}{dt}+\beta P&=&-\frac{\partial U}{\partial X}-qf(t),
\\
\alpha[\dot{l}^2-2l\ddot{l}-2\beta l\dot{l}]&=& \frac{l^2}{l_{0}^2}\left[1+\frac{P^2}{M_{0}^2}\right]-1+\frac{2l^2}{M_{0}l_{0}}\frac{\partial U}{\partial l}.
\label{15b}
\end{eqnarray}
We thus see that indeed, in case the kink width oscillates, it necessarily 
couples to the translational motion, as the derivative of $U(X)$ in Eq.\ 
\ref{15}) contains $l$, whereas $l$ is in turn directly affected by the
momentum. As we observe such oscillations in our simulations, our first
approach with one degree of freedom must necessarily be incomplete.  

The final step is to compare the improved theory to the simulations. 
Lacking analytical solutions, we 
have numerically solved the ordinary differential equations (\ref{12})
and (\ref{13}), computing the kink velocity $\dot{X}(t)$ and its
mean value. 
As shown by Fig.\ \ref{f.2}, the comparison between our improved collective
coordinate theory and the simulations 
is now excellent, as the window numbers 
are correctly estimated and their locations are very accurately predicted. 
We thus see that although the point particle approximation (collective
coordinate $X(t)$) is enough to predict the appearance of a ratchet
phenomenon, the correct description of the 
dynamics necessitates one additional degree
of freedom, $l(t)$, arising from the fact that the fluxon, the ``particle''
in the ratchet, is an extended object that can show internal oscillations.
Strikingly, 
in this way, the interplay of the two degrees of freedom leads eventually
to a behavior truly indistinguishable from a point-particle ratchet. 

\section{Conclusions}

In conclusion, we have proposed and tested in simulations
an experimentally feasible
and uncomplicated procedure to build a soliton ratchet using modified
long Josephson junctions. The main advantage of this system is its 
simple design, that allows an easy implementation by means of 
indentations of the insulating layer (microshorts). 
 An interesting feature of our system is its ability to rectify 
ac forces even if they are time-symmetric, something that cannot 
be accomplished by the homogeneous sG model \cite{flach,salerno}.
We have been able to show analytically that the physical mechanism
responsible for the appearance of the ratchet effect is the coupled
dynamics of the center and width degrees of freedom, whose combined
evolution is able to make fluxons (extended objects) behave as
point-like particles. 
It is important 
to remark that the reported phenomenon is robust as we have 
checked that the ratchet effect survives even in the presence of
noise \cite{futuro}. 

Finally, as regards the generality of the procedure presented 
here, we want to stress that our
results open the door to many other applications. Indeed, 
the mechanism for the ratchet effect we have found, namely 
the coupling between translational motion and internal oscillations,  
will be relevant in general for topological solitons, 
such as those found in the
$\phi^4$ and other nonlinear Klein-Gordon models. Such models 
describe 
applications in a variety of fields 
ranging from biophysics to pattern forming systems (see, e.g.,
\cite{angelbis,Ustinov} for references). Work along these
lines, oriented specifically to assess the actual role of this
phenomenon related to the macromolecules modelled by the $\phi^4$ 
equation (see \cite{moco} and references therein), is in progress.

\acknowledgments
This work has been
supported by the Ministerio de Ciencia y Tecnolog\'\i a of Spain
through grant BFM2000-0006 (AS),
and by the International Research Training Group
`Nonequilibrium Phenomena and Phase Transitions in Complex Systems'
(DFG, Germany).

\end{document}